\documentclass[12pt]{article}
\usepackage{epsfig}
\usepackage{amsmath,amssymb}
\usepackage[usenames]{color}
\def\beq{\begin{equation}}
\def\eeq#1{\label{#1}\end{equation}}
\def\eeqn{\end{equation}}

\def\beqa{\begin{eqnarray}}
\def\eeqa#1{\label{#1}\end{eqnarray}}
\def\eeqan{\end{eqnarray}}

\let\bar=\overbar

\def\Dslash{\not{\hbox{\kern-4pt $D$}}}
\def\dslash{\not{\hbox{\kern-2pt $\del$}}}

\def\msb{{\bar{\ssstyle M \kern -1pt S}}}

\usepackage{fancyhdr,graphicx}
\def\Title#1{\begin{center} {\Large {\bf #1} } \end{center}}

\begin{document}

\Title{Stability windows at finite temperature}

\bigskip\bigskip


\begin{raggedright}

{\it J. R. Torres$^\dagger$\index{Vader, D.}, \it D. P. Menezes$^\dagger$\index{Vader, D.}, \it V. Dexheimer\index{Vader, D.}$^\ddagger$,\\
$^\dagger$Departamento de F\'isica\\
Universidade Federal de Santa Catarina\\
88040-900 Campus Universit\'ario\\
Florian\'opolis, SC\\
Brazil\\
$^\ddagger$Physics department, Gettysburg College, Gettysburg, USA\\
{\tt Email: james.r.torres@posgrad.ufsc.br}}
\bigskip\bigskip
\end{raggedright}

\section{Introduction}

The assumption underlying the existence of quark stars is based on the Bodmer-Witten conjecture \cite{bodmer_witten}. These authors have claimed that it is possible that the interior of a neutron-like 
star does not consist primarily of hadrons, but rather of the strange matter (SM). Strange matter is composed of deconfined quarks, including up, down and strange quarks, plus the leptons necessary to ensure charge neutrality  and $\beta$-equilibrium. 
This possibility arises because a phase transition from hadronic to quark phase is possible at densities present in the interior of neutron stars. It has been argued \cite{olinto} that strange matter is the 
true ground state of all matter. If this is the case,  as soon as the core of the star converts to the quark phase, the entire star converts. SM was first considered in calculations obtained within the MIT
bag model framework \cite{mit1_mit2}. More sophisticated treatments for SM, based on the quark-mass density dependent model \cite{chakrabarty91,lugones95,peng00} is considered in this work.
We discuss next the following models for proto-quark stars: the QMDD model \cite{lugones95}, and the MIT bag model \cite{mit1_mit2}. With QMDD model, the effective quark masses, read

\begin{equation}
m_{u,\overline{u}}^{\ast }=m_{d,\overline{d}}^{\ast }=\frac{C}{3n_{B}},~~~~ m_{s,\overline{s}}^{\ast }=m_{s,\overline{s}}+\frac{C}{3n_{B}} \mbox{,}
\label{ansatz_massa}
\end{equation}
where $C$ is the constant energy density in the zero quark density limit and baryonic density $n_{B}=1/3\left(\rho_{u}+\rho_{u}+\rho_{d}\right)$.
The pressure, energy density and baryonic density of the system are respectively given by

\begin{equation*}
\begin{split}
p&=\sum_{i}\frac{\gamma_{i}}{3}\int \frac{d^{3}k_i}{\left(2\pi \right)^3}\frac{k_i^{2}}{\sqrt{k_i^{2}
+m_{i}^{\ast2}}}\left(  {f_{+ i}}+{f_{- i}} \right)-\mathcal{B}(n_{B},f_{\mp i}) \mbox{,}\\
\end{split}
\end{equation*}
\begin{equation}
\begin{split}
\epsilon&=\sum_{i}\gamma_{i}\int\frac{d^{3}k_i}{\left(2\pi \right)^3}\sqrt{k_i^{2}+m_{i}^{\ast2}}\left({f_{+ i}}+{f{_-,i}} \right)+
\mathcal{B}(n_{B},f_{\mp i}) \mbox{,}\\
\rho_B&=\sum_i \frac{\rho_{i}}{3}=\sum_i \frac{\gamma_{i}}{3}\int \frac{d^{3}k}{\left(2\pi \right)^3}\left({f_{+,i}}-{f_{-,i}} \right) \mbox{,}\\
\mathcal{B}(n_{B},f_{\mp})&=\sum_{i}\gamma_{i}\int \frac{d^{3}k_i}{\left(2\pi \right)^3}\frac{m_{i}^{\ast}}{\sqrt{k_i^{2}+m_{i}^{\ast2}}}
\times\left( \frac{C}{3n_{B}}\right)\left({f_{+ i}}+{f_{- i}} \right) \mbox{,}
\end{split}
\end{equation}
and $\gamma_i$ is the degeneracy of each quark $i=u,d,s$ taking into account spin and number of colors. The distribution function for quarks ($f_+$) and 
antiquarks ($f_-$) are the Fermi-Dirac distributions
$f_{\pm i}=\left[ 1+\exp \left[ \left( E_{i}^{\ast } \mp \mu _{i}\right)/T\right] \right]^{-1}$ where $\mu_i$ is the chemical potential of 
each particle species and $E_{i}^{\ast }\left( p\right) =\sqrt{k_i^{2}+m_{i}^{\ast 2}}$. For the MIT bag model the EoS are obtained as above, but with a simple modification:
the quark masses are fixed, $m^{*}$ is replace by $m$ and $\mathcal{B}(n_B,f_{\pm})$ by a bag constant $\mathcal{B}$.
We take advantage of the simplicity of the MIT bag model
  and include magnetic field effects in the calculation as done in
  Ref.~\cite{veronica}.

\section{Stability windows}
  An important ingredient in the SM hypothesis is the stability windows (SW), identified with the model parameters that are consistent with the fact that two-flavor quark
matter must be unstable (i.e., its energy per baryon has to be larger than 930 MeV, the lead binding energy) and SM (three-flavor quark matter) must be stable (i.e., its
energy per baryon must be lower than 930 MeV).
At zero temperature, the stability window is normally obtained for neutral matter in $\beta$-equilibrium, as for instance 
in Ref.~\cite{lugones95,1995NuPhA.588..365T,Prakash:1996xs}. Nevertheless, one has to bear in mind that stable nuclear matter (as in iron) 
is not charge neutral and does not contain electrons. Actually, its proton fraction is 0.46, very close to symmetric matter and this 
is a good reason to analyze also matter with equal quark chemical potentials, as done in Ref.~\cite{james} for zero temperature systems.
We start by analyzing the stability window related to proto-quark stars described by SM 
equations of state. We have used, for two-flavor quark matter (2QM), the fact that $\mu_u=\mu_d$, which gives symmetric matter ($\rho_u=\rho_d$) and, to be consistent, for SM we have used $\mu_u=\mu_d=\mu_s$. 
Moreover, instead of considering the binding energy \cite{chmaj89,lugones95t}, the quantity that has to be studied is the free energy.
 As expected from calculations in the macrocanonical or grand-canonical ensemble, the quantity related to the thermodynamical potential 
is the free energy per baryon (${\cal F}/A = (\epsilon - Ts)/\rho_B$), where $f$ is the free energy density, $\rho_B$ the baryon density, 
$\epsilon$ the energy density, $T$ the temperature and $s$ the entropy density of the system \cite{greiner}. For zero temperature systems, the free 
energy density becomes the energy density and hence, the binding
energy per baryon ($B/A=\epsilon/\rho_B$) is the correct quantity to be
analyzed at zero temperature in the search for stable 
matter. The choice of appropriate parameters compatible with stable SM at finite temperature systems requires, in addition to the 
investigation performed at zero temperature, a careful study of the
free energy per baryon at finite temperature \cite{veronica,wen2005}.

\begin{figure}
\center
\includegraphics[width=0.55\textwidth]{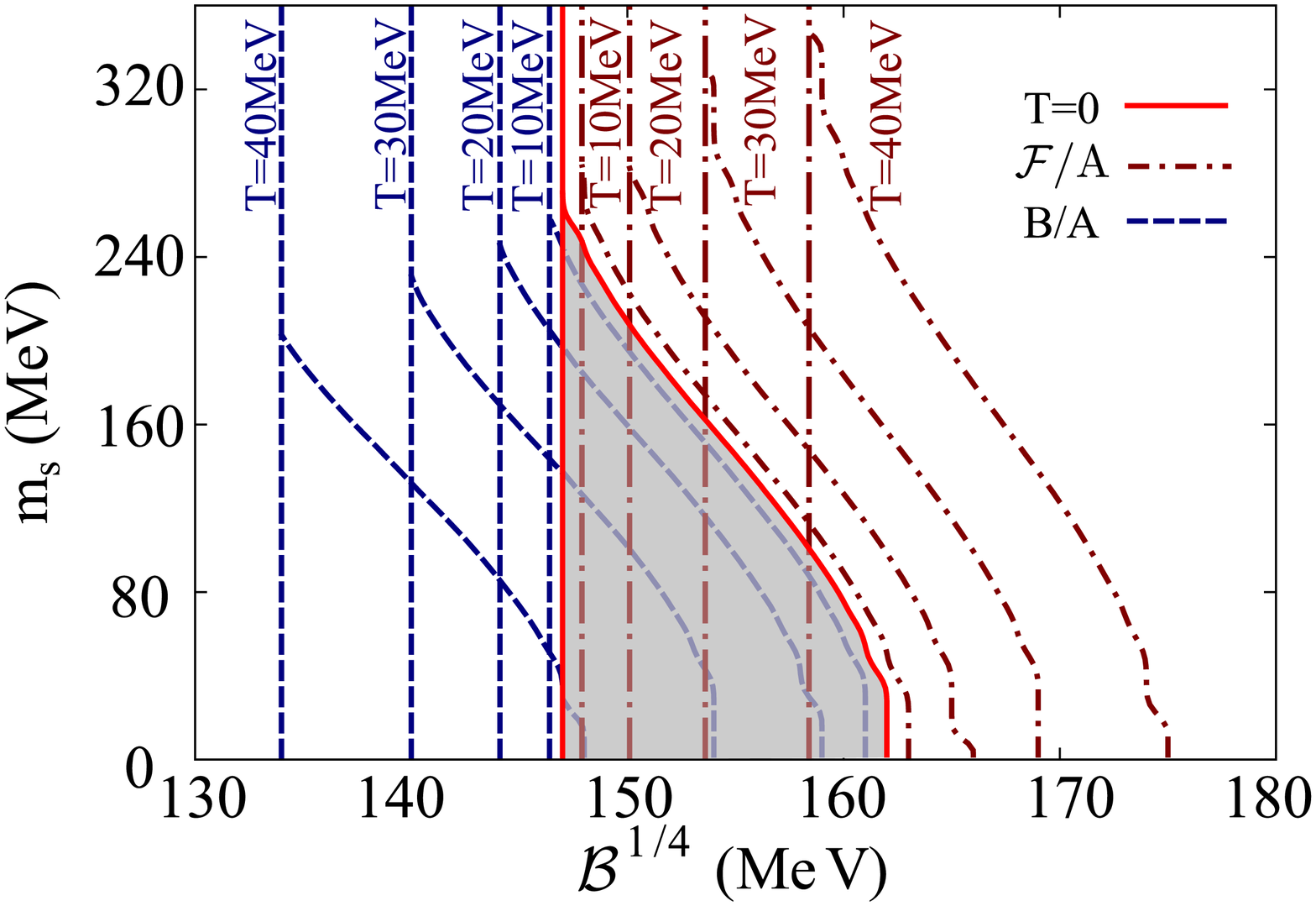}
\includegraphics[width=0.45\textwidth]{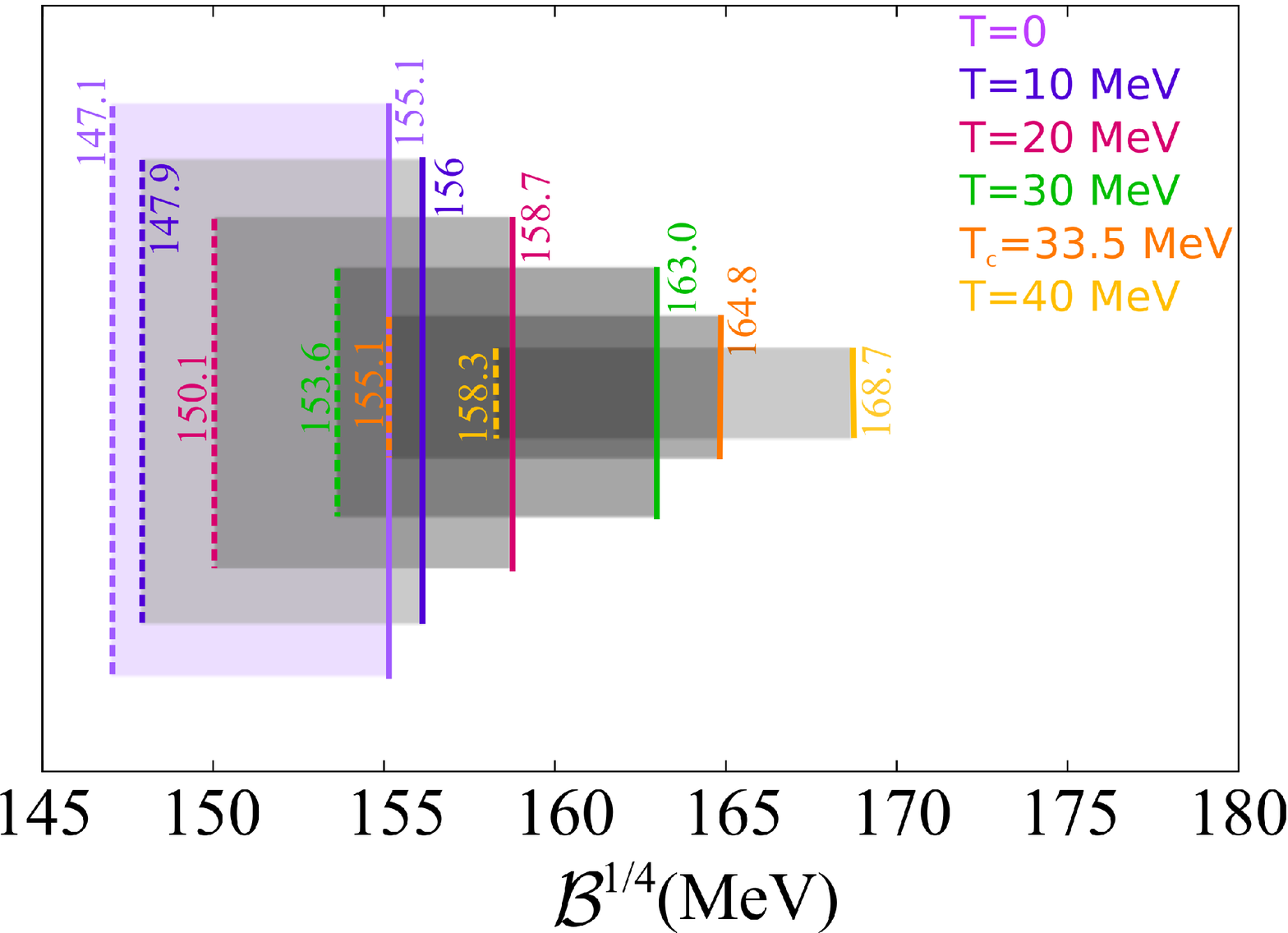}
\includegraphics[width=0.45\textwidth]{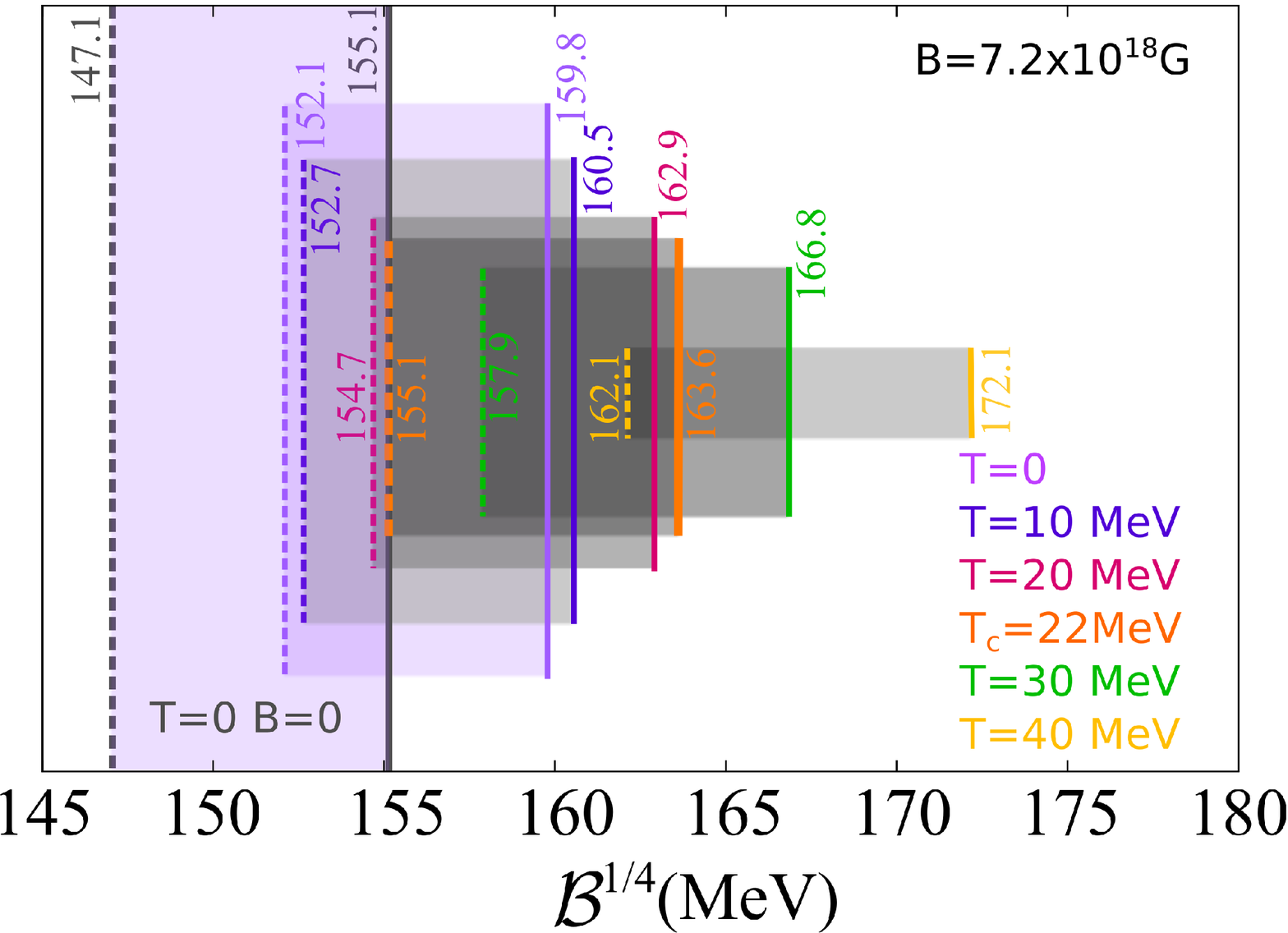}
{\textbf{Fig.01} SW of SM  obtained with the MIT model (up) through the analysis of the free energy per baryon and the binding energy 
per baryon and (bottom) and with $m_{s}=150$ MeV without (left) and with (right) the inclusion of a
magnetic field of $7.2\times10^{18}$ G for different temperatures.}
\label{1}
\end{figure}

\begin{figure}
\centering
\includegraphics[width=0.45\textwidth]{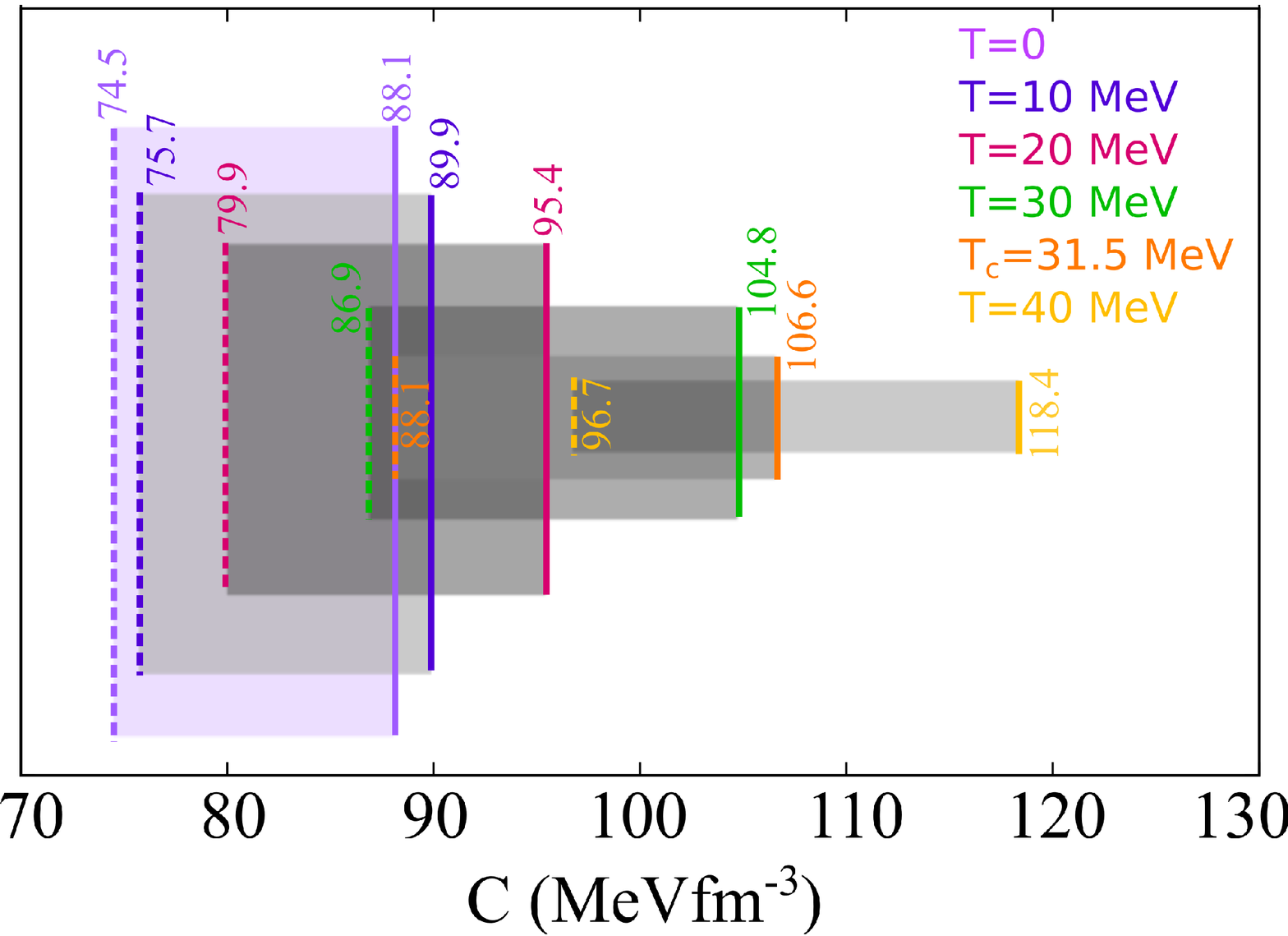}
\includegraphics[width=0.45\textwidth]{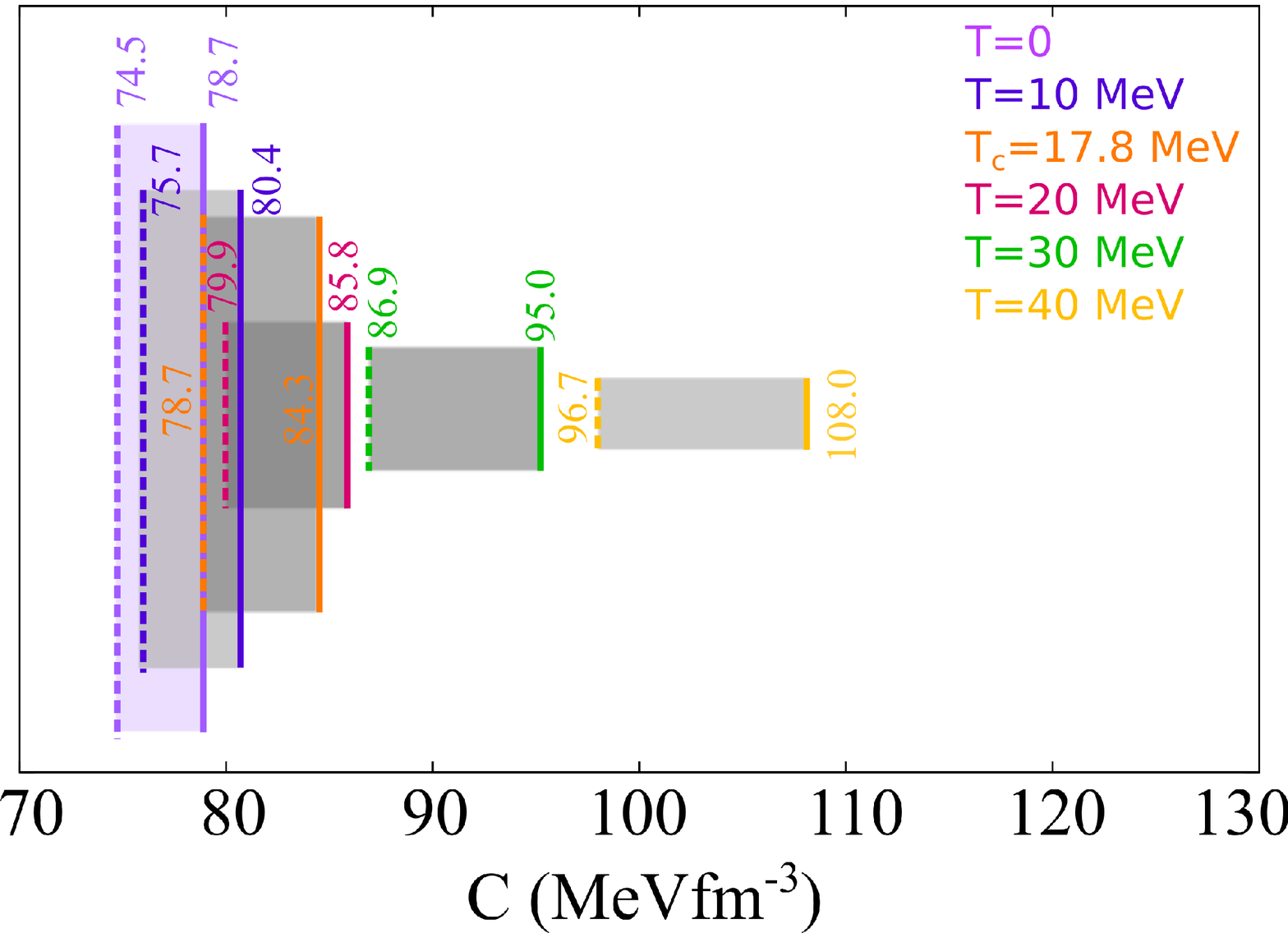}
{\textbf{Fig.02} SW of SM obtained with the QMDD model through the analysis of the free energy per baryon shown for different temperatures and 
$m_{0s}=100$ MeV (left) and (right) with $m_{0s}=150$ MeV.}
\end{figure}

\section{Results}

 In Fig.01 (up) we map in the parameter space of the MIT bag model the
  values that satisfy the stability conditions considering both the
  binding energy and the free energy density. 
The lower limit, vertical straight line, is due to the requirement
that 2QM is not absolutely stable. We have repeated the calculation for temperatures up to 40 MeV, as they 
are relevant for the proto-neutron star evolution simulations.
The shaded area corresponds to the zero temperature results.
For the QMDD we obtain a very similar trend \cite{veronica}.
Our results reproduce the same behavior as the ones given in Refs.~\cite{chmaj89,lugones95t} for the binding 
energy per baryon, i.e., as temperature increases, the stability
windows move to the left, i.e., becomes more restrictive.
However, when we consider the free energy per baryon, the stability
window moves to the right and a wider range of constants becomes
possible to ensure stable matter. In Fig.01 (bottom) the SW obtained 
with the MIT model without (left) and with (right) the inclusion of a 
magnetic field of $7.2\times10^{18}$ G is shown for different
temperatures and $m_{s}=150$ MeV. Magnetic fields make the range of
possible parameters that satisfy stability conditions wider.
In Fig.02 the SW for the QMDD model with $m_{s}=100$ (left) and
$m_{s}=150$ (right) are shown. A larger $s$ quark mass yields a
smaller range of possible parameters that satisfy the stability conditions.

We presented in this paper a review of stability windows for quark matter
at finite temperature with two models. 
We pointed out that the correct quantity to analyze in order to obtain
the upper limits of the stability windows is the free energy per baryon. For both models under investigation,
the  constants  that are allowed by
stability conditions belong to a wider range as the temperature increases.

{\bf Acknowledgments}:This work was supported by CAPES/CNPq/FAPESC.

\end{document}